\documentclass{article}
\usepackage{arxiv}
\usepackage{color}
\usepackage{cite}
\usepackage{subfigure}
\usepackage{graphicx}
\usepackage{dcolumn}
\usepackage{bm}
\usepackage{braket}
\usepackage{floatrow}
\usepackage{amsmath}
\usepackage{makeidx}
\usepackage{epsfig}
\usepackage{array}
\usepackage{epstopdf}
\usepackage{multirow}
\usepackage[utf8]{inputenc} % allow utf-8 input
\usepackage[T1]{fontenc}    % use 8-bit T1 fonts
\usepackage{hyperref}       % hyperlinks
\usepackage{url}            % simple URL typesetting
\usepackage{booktabs}       % professional-quality tables
\usepackage{amsfonts}       % blackboard math symbols
\usepackage{nicefrac}       % compact symbols for 1/2, etc.
\usepackage{microtype}      % microtypography
\usepackage{lipsum}	
\usepackage{listofsymbols}
\usepackage{moreverb}
\usepackage{listings} %sourcecode	
\title{Machine Learning of Nonequilibrium Phase Transition in an Ising Model on Square Lattice}
\author{Dagne Wordofa \\
    Department of Physics\\
    Addis Ababa University\\
    Addis Ababa, Ethiopia\\
    \texttt{dagnehordofa@gmail.com}\\
\And
 Mulugeta Bekele\thanks{Corresponding author}\\
 Department of Physics\\
 Addis Ababa University\\
 Addis Ababa, Ethiopia\\
 \texttt{mulugetabekele1@gmail.com}
}
\begin{document}
\maketitle
\begin{abstract}
This paper presents the investigation of convolutional neural network (CNN) prediction successfully recognizing the temperature of the  non-equilibrium phases and phase transitions  in two-dimensional (2D) Ising spins on square-lattice.
The model uses image snapshots of ferromagnetic 2D spin configurations as an input shape to provide the average out put predictions.
By considering supervised machine learning techniques,
we  perform the (modified) Metropolis Monte Carlo (MC) simulations to generate the equilibrium (and non-equilibrium) configurations.
In equilibrium Ising model, the Metropolis algorithm respects detailed balance condition (DBC), while its modified non-equilibrium version violates the DBC. Violating the DBC of the algorithm is characterized by a parameter $-8 <  \varepsilon  < 8$.
We find the exact result of the transition temperature in terms of $\varepsilon$. This solution is used to encode the two (high-and low-temperature) phases through an order parameter of the model.
If we set $\varepsilon = 0$, the usual single spin flip algorithm can be restored and the equilibrium configurations (training dataset) generated with such set up are used to train our model.
For $\varepsilon  \neq 0$, the system attains the non-equilibrium steady states (NESS), and the modified algorithm generates NESS configurations (test dataset),  not defined by Boltzmann distribution.
 Finally, the trained model has been  validated and  successfully tested on the test dataset. Our result shows that CNN  can correctly determine the nonequilibrium phase transition temperature $T_c$ for various $\varepsilon$ values, consistent with the exact result (our study) and also in agreement with MC result (literature).
\end{abstract}
\keywords{Nnonequilibrium, Phase Transition, Ising, Critical Temperature,  Machine Learning
}

%%%%%%%%%%%%%%%%%%%%%%%%%%%%%%%%%%%%%%%%%%
%\setcounter{section}{-1} %% Remove this when starting to work on the template.
%%%%%%%%%%%%%%%%%%%%%%%%%%%%%%%%%%%%%%%%%%%%%%%%%%%%%%%%%%%%%%%%%%%%%%%%%%%%%%%%%
\section{Introduction}\label{section:Introduction}
%%%%%%%%%%%%%%%%%%%%%%%%%%%%%%%%%%%%%%%%%%%%%%%%%%%%%%%%%%%%%%%%%%%%%%%%%%%%%%%%%
Currently, the standard theory and general framework for the critical phenomenon near continues  phase transitions (PT) is well understood in equilibrium systems~\cite{Kardar2007, Nishimori2010, Goldenfeld2018}. However, the study of phase transitions between non-equilibrium statistical states have consistently been among the main subjects of ongoing research and exploration~\cite{Derrida2007, Derrida2011, Bertini2015, Godreche2009, Stinchcombe2001,Mukamel2000,Odor2004,Hinrchsen2000}.
Identifying the critical points of various phases within the parameter space is a fundamental undertaking in the fields of statistical and condensed-matter physics.
Machine learning (ML) is the field of study concerned with algorithms that are designed to improve their performance by getting experience from data~\cite{Ethem2004}.
Relatively recently, the utilization  of these techniques have been successfully employed in various domains such as investigating the phases of the Ising model~\cite{Tanaka2017, Walker2018, Alexandrou2020, Burak2022}, phase transition in the Bose-Hubbard~\cite{Huembeli2018,Dong2018}, disordered quantum systems~\cite{Ohtsuki2016,Ohtsuki2019}, and material properties~\cite{Pilania2013}.

In this report, we introduce the application of ML to non-equilibrium PT  which can be accomplished based on the well established features of modern theories of PT in equilibrium systems. In equilibrium systems, PT is generically described by singularities in the free energy and its derivatives. Such singularity causes a discontinuous property of thermodynamic quantities near the transition point. Phenomenologically, the PT is defined regarding to an \emph{order parameter}, which has a zero value in the ordered phase while it vanishes in the disordered phase~\cite{Onsager1944,YangI1952,YangII1952}. Within the scope of this paper, the paradigmatic example that we will be working on is a two-dimensional (2D) Ising spin system on a square lattice. It is interesting to note that the 2D Ising spin on square-lattice is a simple that can be exactly solved~\cite{Onsager1944}. Despite the fact that it is exactly solvable, it is still a topic of ongoing research that is frequently used in the context of ML~\cite{Tanaka2017,Walker2018,Alexandrou2020,Morningstar2018,Walker2020,Francesco2020, Carrasquilla2017,Corte2020, avecedo2021_2, Burak2022,ZhenyLi2019, Burzawa2019}. In this investigation, first we try to perform the graphical solution~(\ref{Eq:8}) of the nonequilibrium transition temperature, see supplementary page~\ref{appendix:A1}. Then we look at the possibility of a non-equilibrium phase transitions occurring within the Ising model that breaks the principle of detailed balance through machine learning. To be more explicit, we aim to find the non-equilibrium phases and  the transition  temperatures by applying convolutional neural networks (CNN) based on the general framework of supervised learning discussed in~\cite{Bahri2020}.
This framework was reviewed before in Statistical Mechanics of deep learning, which also briefly explain  the connection between deep learning and the modern subject of Statistical Physics.

According to the findings presented in Ref.~\cite{Carrasquilla2017}, the application of ML to the issue of phases of matter has, for the most part, been effective, and motivated with this work, we aim to extend this ML application to the case of \emph{non-equilibrium} PT in 2D Ising model. For its compatibility, the Ising model which was addressed in~\cite{Kumar2020} becomes the primary focus of our attention. We employ the Monte Carlo (MC) approach~\cite{berg2004,LandauBinder2014, Metropolis1953, Janke2007} to generate a properly distributed data set of Ising spin configurations on $L \times L$ square lattice (where $L$ is its leaner size), together with their associated labels, while taking supervised learning techniques into consideration.
Accordingly, by the context of equilibrium and non-equilibrium systems, we are refiring to two different spin update rules; (i) rule that holds the DBC and (ii) rule that breaks the DBC, respectively. The former is used to generate the \emph{train-dataset}, while the latter is used to generate the \emph{test-dataset} which can be seen from some representative configurations illustrated in~\ref{appendix:A2} (Figure~\ref{fig:6}).

We build a CNN using open-source software~\cite{TensorFlow2015}, see supplementary information~(\ref{appendix:A3}) and an example shown in Figure~\ref{fig:cnn1} for  more details.
We train our model on the train-dataset and it has been effectively validated to classify simulation results of the equilibrium 2D Ising model into the ferromagnetic (FM) or "ordered state" and the paramagnetic (PM) or "disordered state" phases. The classification of these results was also successfully validated, for example~\cite{Carrasquilla2017, Corte2020}. The main goal of this work is to evaluate the generalization reach of the CNN by testing it with configurations (test-dataset) from a system that is not in equilibrium. Intriguingly, in addition to accurately categorizing the configurations, we will demonstrate that the CNN can exhibit the critical temperature of the non-equilibrium PT. Our findings is very close to the exact solution~(\ref{Eq:8}), and also  consistent with the MC results provided in Ref.~\cite{Kumar2020}.

The remaining sections are organized as follows: Next, we present the model considered in this research, followed by a concise description of the Metropolis MC method for generating image samples of Ising configurations in Section~\ref{section:Model}.
Some of the results of this study are then illustrated in Sec.~\ref{section:Results}. Finally, we provide a summary of the main results and discussion as presented in Sec.~\ref{section:Conclusion}.

%%%%%%%%%%%%%%%%%%%%%%%%%%%%%%%%%%%%%%%%%%%%%%%%%%%%%%%%%%%%%%%%%%%%%%%%%%%%%%%%%
\section{Description of the Model and Metropolis Monte Carlo Method}\label{section:Model}
%%%%%%%%%%%%%%%%%%%%%%%%%%%%%%%%%%%%%%%%%%%%%%%%%%%%%%%%%%%%%%%%%%%%%%%%%%%%%%%%%
We consider the  2D Ising model on a square lattice of linear size $L$ sites.
 The system size ($\mathcal{N}= L \times L$) is equal to the total number  of  spins ($N$), which means that each of the site contains one spin that points either up or down ($\pm 1$). If we assume zero magnetic field, the nearest-neighbor interaction energy of (ferromagnetic) Ising model is given as,
\begin{equation}\label{Eq:1}
   E = - J \sum_{\langle i, j  \rangle}\sigma_{i} \sigma_{j},
\end{equation}
 where $\sigma_i = \pm 1$ denotes the value of the spin at site $i= \{1,\cdots, \mathcal{N}\}$, the indices $\langle i,j \rangle$ represent the nearest-neighbor pairs~\cite{berg2004,LandauBinder2014}, and a ferromagnetic energy scale $J>0$ refers to the strength of exchange interaction.

At the critical (or transition) temperature, the system  exhibits a second order phase transposition.
The transition temperature of the nearest-neighbor equilibrium Ising model, for an infinite square lattice, was derived~\cite{Onsager1944} to be $ 2/\ln(1+\sqrt{2})$, see Eq. (\ref{Eq:6}).
In this case, the system is assumed as a {\em magnetized state} when its temperature is lower than $ 2/\ln(1+\sqrt{2})$, which is known as the ordered state (FM phase). On the other hand, the system is said to be in the disordered state (PM phase) if its temperature is higher than $ 2/\ln(1+\sqrt{2})$. The  magnetization per spin is what determines the value of the order parameter,
\begin{equation}\label{Eq:2}
m = \frac{1}{N} \left| \sum_i^{N} \sigma_i \right|.
\end{equation}
This quantity~(\ref{Eq:2}) distinguishes the \emph{two phases} that are realized by the system. It is zero(nonzero) in the disordered(ordered) phase.

\subsection{The Modified Metropolis Algorithm}
Let us consider a system that is in contact with a heat bath and produces stochastic spin flips, following Ref.~\cite{Glauber1963}.
In the context of the equilibrium Ising model, it can be observed that the system attains thermal equilibrium over a significant time and thus the
steady state distribution can be accurately described by the Boltzmann distribution. This is a valuable approach for establishing transition rates and calculating the probabilities of spin flipping. The Metropolis algorithm~\cite{Metropolis1953} is the transition rate that is commonly used and can be stated as
\begin{equation}
W =  \texttt{MIN} \left[1,  \texttt{e}^{-\Delta E/k_{B}T}\right],
\label{Eq:Metropolis}
\end{equation}
where $W$ represents the rate of change from  state \texttt{b}(\emph{before} flip) to another state \texttt{a}(\emph{after} flip),
$ \Delta E = E_{\texttt{a}} - E_{\texttt{b}} $ is the change in energy that occurs as a result of this transition, and  $k_{B}$ denotes the known Boltzmann's constant. In this context, the unit of temperature $T$ is linked to the units of $J/k_{B}$. (For the remainder of this description, we will assume $k_B= 1$, thus $ T = T/J$ becomes dimensionless.)
The defined algorithm~(\ref{Eq:Metropolis}) meets the requirements of the detailed balance condition (DBC). The aforementioned statement denotes that there exists a microscopic reversibility of every elementary process, which is counterbalanced by its corresponding reverse process  \cite{Zia2007}.
That is $
W_{\texttt{b}\rightarrow\texttt{a}}p_{\texttt{eq}}^{\texttt{b}} =  W_{\texttt{a}\rightarrow\texttt{b}}p_{\texttt{eq}}^{\texttt{a}}$,
where $p_{\texttt{eq}}^{\texttt{b}} \propto  \exp[-E_{\texttt{b}}/T]$.
Therefore the ratio $\texttt{w}=W_{\texttt{b}\rightarrow\texttt{a}}/W_{\texttt{a}\rightarrow\texttt{b}}$ gives $\texttt{w} = \exp[-\Delta E/T]$.

\begin{figure}[hbpt]
     \centering
 \includegraphics[width=0.45 \columnwidth]{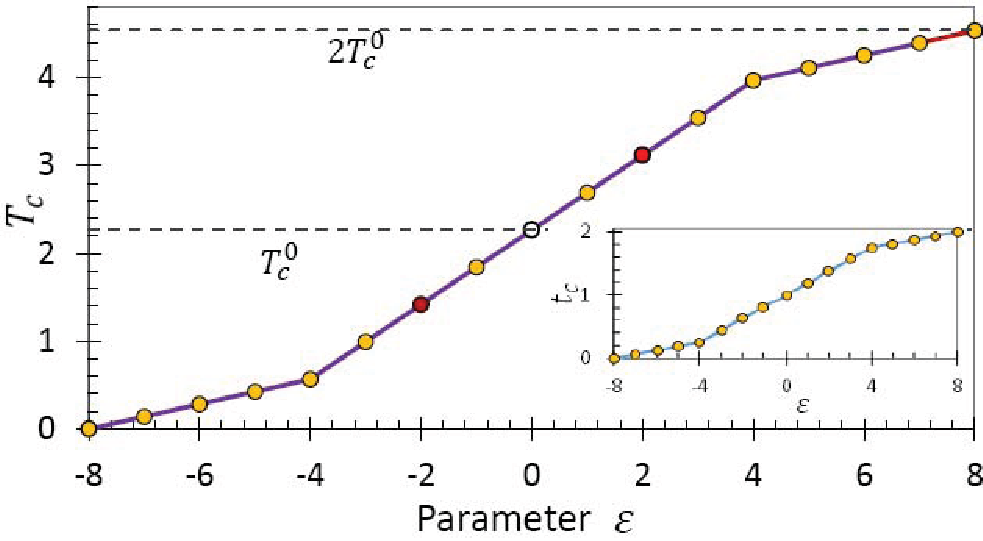}   %Tcab2.png}
    \caption{\small{Transition temperature $T_{c}$  as a function of parameter $\varepsilon$. A Plot of Eq. (\ref{Eq:8}).
 The horizontal dashed lines represents $T_{c}= T_{c}^{0}$ and  $T_{c}= 2T_{c}^{0}$ as shown where  $T_{c}^{0} = 2/\ln(1+\sqrt{2})$.
The inset shows $t_{c}$ versus $\varepsilon$ where $t_{c} = T_{c}/T_{c}^{0}$.}} \label{fig:1}
\end{figure}

The topic of non-equilibrium phase transitions is examined with emphasis on fundamental characteristics such as the role of DBC violation in generating effective (long-range) interactions~\cite{Bahri2020}. The equilibration process is not solely dependent on the presence of DBC, as it serves as a sufficient but not a necessary condition. The objective of this study is to deliberately violate the DBC in order to induce a state of \emph{fluctuation} in the system. As noted in reference~\cite{Kumar2020}, there exists a scenario in which the system undergoes an order-disorder phase transitions that different from the typical transitions of the equilibrium case.
If  $ \varepsilon \neq 0 $ denotes the parameter violating the DBC, it is possible to substitute  $\Delta E$ in Eq.~(\ref{Eq:Metropolis}) with
\begin{equation}\label{Eq:dEeff}
\Delta E_\texttt{eff} = \Delta E + \varepsilon,
\end{equation}
 and the ratio becomes $\texttt{w} =  \texttt{e}^{-\beta( \Delta E + \varepsilon)}$.
It can be inferred that when $\varepsilon$ is positive, $\Delta E_\texttt{eff}$ is greater than $\Delta E$, whereas when $\varepsilon$ is negative, $\Delta E_\texttt{eff}$ is less than $\Delta E$. The former does not facilitate the process of spin flipping, whereas the latter significantly promotes the likelihood of spin flipping. In contrast to spins subjected to the conventional Metropolis algorithm~(\ref{Eq:Metropolis}), spins subjected to the modified flipping rates effectively undergo distinct (transition) temperatures. When  $ \varepsilon < 0$( $ \varepsilon > 0$), it is reasonable to assume that the spins are coupled to a reservoir at a higher (lower) effective temperature ($T_{\texttt{eff}}$). It should be noted that $T_{\texttt{eff}}$ is not uniform across all spins in the system, see appendix  \ref{appendix:A1}.  Thus, ``the system is out-of-equilibrium, and a transition is a non-equilibrium phase transition.
The property of this transition would be a characteristic of the \emph{non-equilibrium steady statey} (NESS) exhibited by the system''~\cite{Kumar2020}. It can be inferred that, unlike an equilibrium system, the distribution of microstates in the NESS cannot be characterized by the Boltzmann distribution.
The transition rate for flipping a spin  $\sigma_{i}^{\texttt{b}}\rightarrow \sigma_{i}^{\texttt{a}} $  can be determined using this definition (\ref{Eq:dEeff}),
\begin{equation}\label{Eq:5}
 W(\pm\sigma_{i} \rightarrow \mp \sigma_{i}) =					\left\{
  \begin{array}{ll}
    \texttt{e}^{-\beta(\varepsilon \pm \Delta E)}, & \hbox{if \; \; }\varepsilon \pm \Delta E  > 0  ; \\
    1, & \hbox{otherwise.}
  \end{array}
\right.
\end{equation}
Here $
 \Delta E = 2 J \sigma_{i}\sum_{j} \sigma_{ij} \equiv \{-8,-4, 0, 8, 4\}[J]$
where $\sigma_{ij}$ refers to  $j=\{\texttt{left, right, top, bottom}\}$ nearest neighbors of the $i^{\texttt{th}}$ \texttt{site}, and the symbol
`$\equiv $' refers to an alternative approach for Ising on a square lattice involves that $ \Delta E $ can assume discrete values from $\{-8, -4, 0, 4, 8 \}$ using the units of $J$. The algorithm given in Eq. (\ref{Eq:5}) still \emph{respects} the DBC when $|\varepsilon| \geq 8$ \cite{EndNote}.
Intriguingly, however, algorithm (\ref{Eq:5}) \emph{violates} the DBC for $-8 < \varepsilon < 8$ (with $\varepsilon\neq 0$) since it is impractical to obtain a unique $T_{\texttt{eff}}$ value for which the transition probabilities for all  feasible $\Delta E$ values that obey the DBC.
According to the established notation, the non-equilibrium phase transitions may occur within the system and the transition temperature $T_c$ must fulfil the relation,
\begin{equation}\label{Eq:6}
T_c = \left\{
  \begin{array}{ll}
   0 < T_{c} <  T_{c}^{0}     & \hbox{if \;  } -8 < \varepsilon < 0; \\
  T_{c}^{0} < T_{c} < 2 T_{c}^{0} & \hbox{if \; \;   }  0 < \varepsilon < 8,
  \end{array}
\right.
\end{equation}
where $T_{c}^{0}  = 2/\ln(1+\sqrt{2}) \approx 2.2692$ is the transition temperature of the \emph{equilibrium} ($\varepsilon = 0$) case.
Explicitly,  we are essentially interested in some  $\varepsilon$ values of  $- 8 < \varepsilon < 8$, as shown in Figure \ref{fig:1}.
Referring to a systematic  graphical solution presented in (\ref{appendix:A1}) the exact result  follows that,
 \begin{align}\label{Eq:8}
 T_{c}^{\texttt{exact}}(\varepsilon) \equiv T_c =   \left\{
   \begin{array}{lcl}
     (0.5 + \varepsilon/16)T_{c}^{0} , & \hbox{for} & -8 < \varepsilon <-4 \hbox{;} \\
     (1 + 3 \varepsilon/16)T_{c}^{0}, & \hbox{>>} & -4 \leq \varepsilon \leq 4 \hbox{;} \\
     (1.5 + \varepsilon/16)T_{c}^{0}, & \hbox{>>} & 4 < \varepsilon < 8 \hbox{,}
   \end{array}
 \right.
\end{align}
where $ T_{c}^{0} \equiv T_{c}(\varepsilon = 0) = 2/\ln(1+\sqrt{2}) $, see Figure \ref{fig:1}.
Specifically, if we focus on  $-4 \leq \varepsilon \leq 4$ that   Eq. (\ref{Eq:8}) can be efficient to discuss the nonequilibrium phase transition.
More specifically, consider two  $\varepsilon$ values ($\varepsilon = \pm2$), conveniently we get that
$T_{c}^{\texttt{exact}}(\varepsilon = \pm2) \approx 3.1201 $($\approx 1.4182 $).
Remarkably, we see that our numerical result (Figure \ref{fig:Result2}) is very close to this result.
%%%%%%%%%%%%%%%%%%%%%%%%%%%%%%%%%%%%%%%%%%%%%%%%%%%%%%%%%%%%%%%%%%%%%%%%%%%%%%%%%
\subsection{Generating 2D Images of Ising Spin Configurations}
Make use of the modified Metropolis rule (\ref{Eq:5}), we achieve Monte Carlo (MC) simulations of the Ising model, see the flow chart shown in~\ref{appendix:A2} (Figure~\ref{fig:9}). The simulations are performed on a square lattice ($\texttt{Lx} = \texttt{Ly}$) of system size $\mathcal{N} =L^{2}$, inducing periodic boundary condition in ($x$ and $y$) directions.
%The total number of spins in the system is $N = L^{2}$, where we consider $L = \{10, 20, 40,  60\}$.
For each system, we start the simulations from an initial, high temperature (with random spin initial configurations) and perform a standard MC sweeps (MCS) for generating the required samples of $L \times L$ Ising spin configurations as data for the supervised ML approach ~\cite{Metropolis1953,berg2004,LandauBinder2014,Janke2007}. Examples of configurations are shown in Figure \ref{fig:samples}.
For all datasets used in Sec. \ref{section:Results}, the simulation was performed with three  $\varepsilon=\{0, 2, -2$ values. First we set $\varepsilon = 0$ and generate the configurations for the train-dataset. This comprises about 80\% of the total data (where 10\% is again reserved for validation).
Next we set $\varepsilon = 2$ to generate the test-dataset which incorporates the remaining 20\% of the total data, and the procedure is the same for $\varepsilon = -2$.
We restart and repeat this procedure for all system sizes.
Furthermore, one can save the trained sequential model using $\texttt{TenserFlow's Keras API}$, and later it can be loaded to test the configurations from different discrete  $\varepsilon$ values. Efficiently, this can be used  to study the qualitative dependence of $T_c$ on the parameter $\varepsilon$, e.g., see \ref{appendix:A4}.
%%%%%%%%%%%%%%%%%%%%%%%%%%%%%%%%%%%%%%%%%%%%%%%%%%%%%%%%%%%%%%%%%%%%%%%%%%%%%%%%%\newpage
  \section{Results}\label{section:Results}
\begin{figure}[hbpt]
     \centering
\includegraphics[width=0.95 \columnwidth]{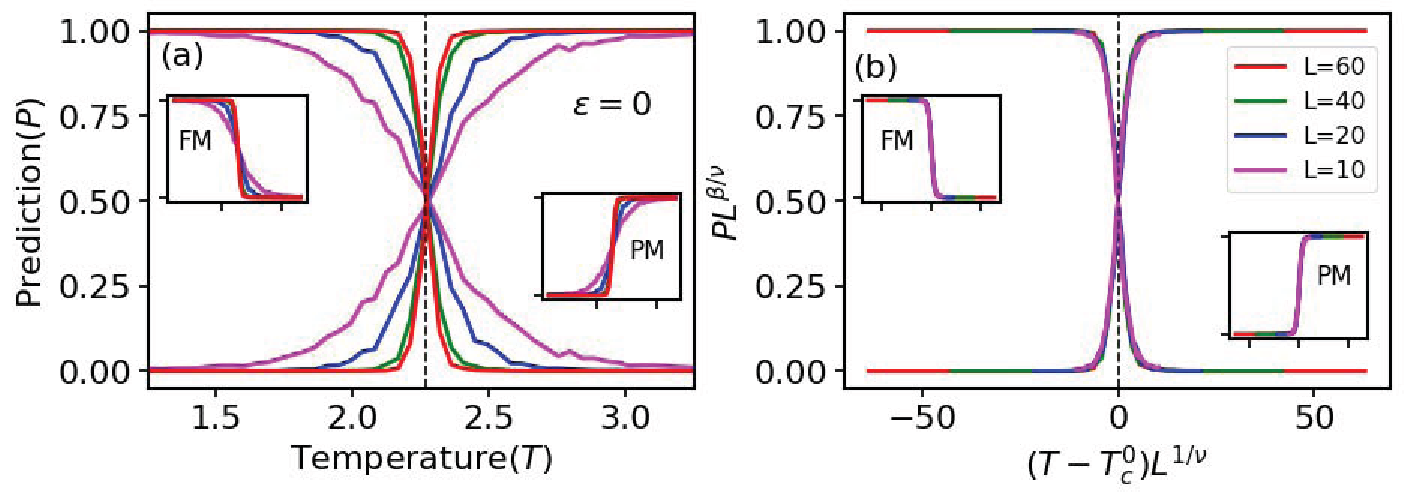}
   \caption{\small{Machine learning (ML) the \emph{equilibrium} ($\varepsilon = 0$) ferromagnetic Ising spin on square-lattice (linear sizes $L= 10, \; 20, \; 40 $ and $60$). (a) The prediction $P$ versus temperature $T$ where the vertical dashed line denotes the estimated value of $T_{c}^{0} \simeq 2.2687 \pm 0.0015 $ of the model. (b) A plot showing data collapse of $PL^{\beta/\nu}$ versus $(T-T_{c}^{0})L^{1/\nu}$. The insets represent FM corves and PM corves as shown.}} \label{fig:Result1}
\end{figure}

%%%%%%%%%%%%%%%%%%%%%%%%%%%%%%%%%%%%%%%%%%%%%%%%%%%%%%%%%%%%%%%%%%%%%%%%%%%%%%%%%
In the current section (Sec. \ref{section:Results}), we briefly present the main numerical results obtained using neural network model (CNN).  Similar to the previous works (literatures), we train the model on equilibrium Ising spin configuration.
After training on an adequately large sample size at temperatures $T > T_c^{0}$ and $T < T_c^{0}$, the CNN can correctly classify configurations in a valid dataset, as illustrated in Figure \ref{fig:Result1}(a) for configurations with the given linear size, $L = \{10, 20, 40, 60\}$.
Systematically, finite-size scaling (FSS) is capable of narrowing in on the thermodynamic result  of  $T_c^{0}$ in a manner comparable to that of magnetization~\cite{Carrasquilla2017}, Figure \ref{fig:Result1}b displays that a data collapse yields a critical exponents estimate of $\nu \approx 1.00 \pm 0.01$ and $\beta \approx 0.125  \pm 0.002$, while a size scaling of the crossing temperature  yields an estimate of $T_{c}^{0} \simeq 2.269$ (see \ref{appendix:A6}).

\begin{figure}[hbpt]
     \centering
\includegraphics[width=0.95 \columnwidth]{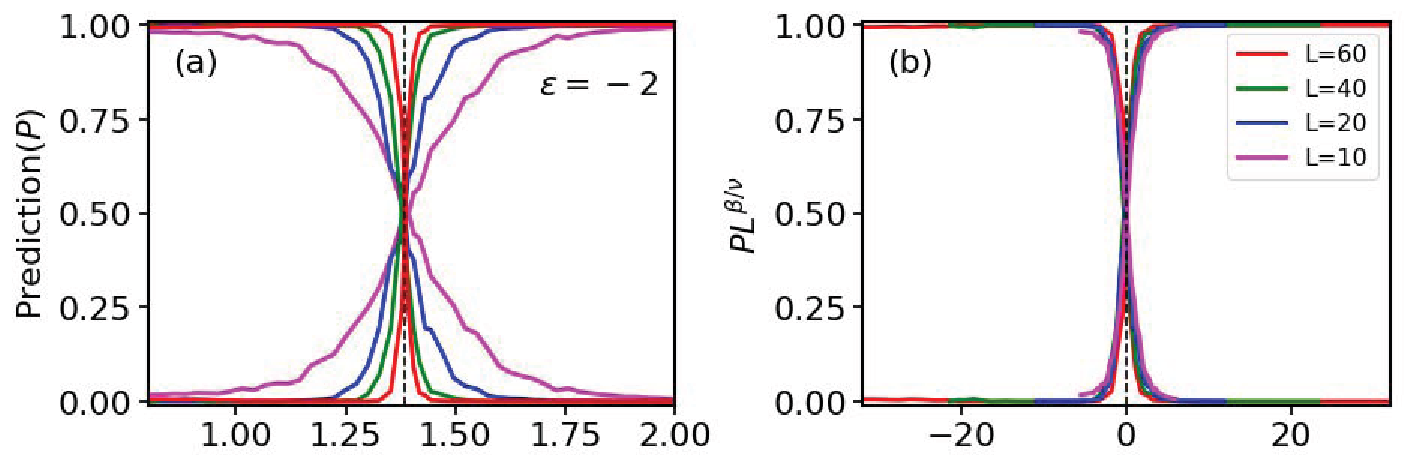}
\includegraphics[width=0.95 \columnwidth]{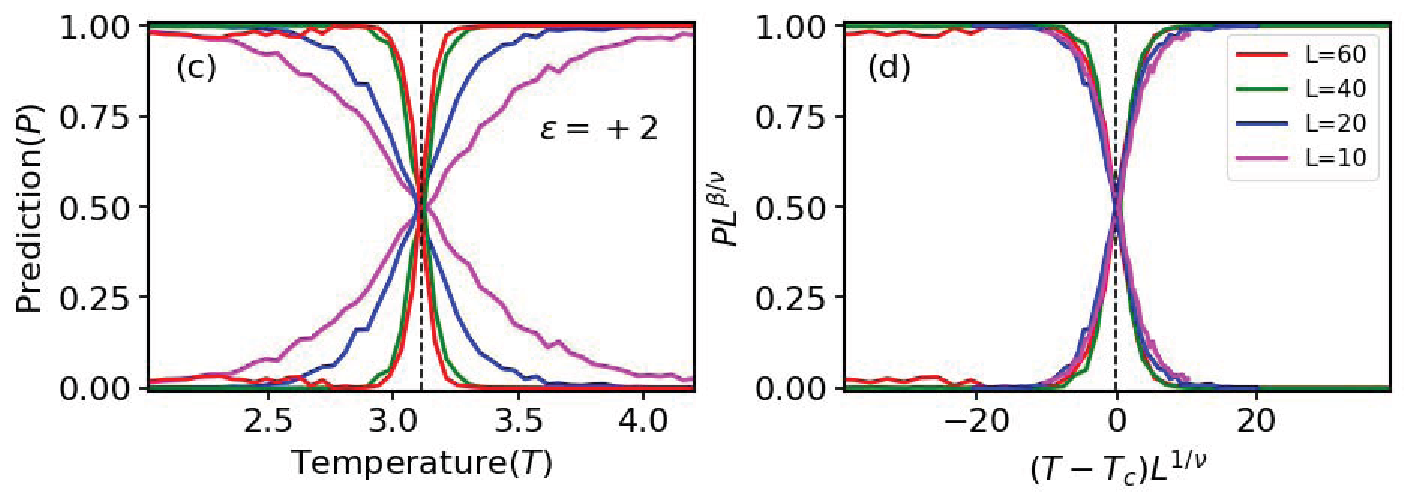}
   \caption{\small{ML the non-equilibrium ($\varepsilon \neq 0$) ferromagnetic Ising spin on square-lattice ($L= 10, \; 20, \; 40 $ and $60$) where $\varepsilon = - 2$ (a) and $\varepsilon = +2$ (c). The left panel (a and c) display $P$ versus $T$ while the right panel (b and d)  represent the corresponding data collapse  $PL^{\beta/\nu}$ versus $(T-T_{c})L^{1/\nu}$.
   The estimated values are indicated by  dashed lines:  $T_{c} \simeq 1.3769\pm 0.0087$  (a), and $T_{c} \simeq 3.1071\pm 0.0175$ (c)}} \label{fig:Result2}
\end{figure}

More interestingly,
``the  generalization competency of the neural networks lies in their ability to provide correct predictions further than the datasets  with which they were trained''.
Accordingly, the trained CNN has been provided with a test dataset of configurations from a 2D Ising model in which data generation was incorporated by changing the update rules where violation of the DBC is accountable.
This is intended to answer the question \emph{``Does CNN that trained on equilibrium phase transition in Ising model  with detailed balance able to  recognise  the non-equilibrium phase transition?''}
Thus, next we present the results of this scenario by using our CNN, which is already trained and validated on configurations for the square-lattice ferromagnetic Ising model, and provide it a test dataset generated by modified Metropolis MC simulations for the same sizes as $L$s  in Figure \ref{fig:Result1}.
%-----------------------
In Figure \ref{fig:Result2}
we illustrate the average of prediction $P$ versus temperature $T$ for configurations from two different test datasets ($\varepsilon = \pm 2$) of each with four linear sizes (see keys).
The dashed lines denote the estimated values of the transition temperatures (a)  $T_{c} \simeq 1.377 $, and (c)  $T_{c} \simeq 3.107 $.
Clearly we see that $T_{c}^{\texttt{ML}}(\varepsilon = \pm 2)$  is close to $T_{c}^{\texttt{exact}}(\varepsilon = \pm 2)$ obtained in Eq. (\ref{Eq:8}). 
On the right panel, `b' and `d'  represent the corresponding data collapse $PL^{\beta/\nu}$ versus $(T-T_{c})L^{1/\nu}$ allowing us to successfully compute the critical exponents, $\nu \approx 1.02 \pm 0.02$ and $\beta \approx 0.126 \pm 0.003$.
Our results are consistent with the MC result reported in \cite{Kumar2020}.
%%%%%%%%%%%%%%%%%%%%%%%%%%%%%%%%%%%%%%%%%%%%%%%%%%%%%%%%%%%%%%%%%%%%%%%%%%%%%%%%%
  \section{Summary and Conclusions}\label{section:Conclusion}
%%%%%%%%%%%%%%%%%%%%%%%%%%%%%%%%%%%%%%%%%%%%%%%%%%%%%%%%%%%%%%%%%%%%%%%%%%%%%%%%%
In order to categorize the two typical phases of ferromagnetic Ising spins on square-lattice, we used supervised machine learning approaches.
Our findings indicate that advanced ML architectures, such as the fully connected CNN, are able to detect non-equilibrium transition temperature $T_c$ so long as they are properly trained on equilibrium Ising spin configurations.
To train (and validate) the model, we use the training dataset generated by running the MC simulations of 2D ferromagnetic Ising system on square-lattice, where the spin`s update rule is governed by the usual Metropolis algorithm. This update rule is compliant with the DBC.
Then we test the model on simulations of 2D ferromagnetic system (test dataset) and, in this case, the update schem is performed using modified version of the algorithm. That is; the modified update rule violates the DBC.
In the model, violating DBC is designated by a parameter $\epsilon$ that is fixed to take values in the range $-8<  \varepsilon < 8$).
We successfully derived the exact solution of the non-equilibrium transition temperature $T_{c}(\varepsilon)$, Eq. (\ref{Eq:8}).
This solution suggests that only the parameter $\varepsilon$ affects the transition temperature.
For $\epsilon = 0$, the  equilibrium transition $T_{c}^{0}$ can be retrieved.
For $\epsilon \neq 0$, the system reaches the NESS; this state cannot be characterized using the Boltzmann distribution, and
 the numerical results are consistent with the exact solution.
 For instance, for $\epsilon = \{-2, 2\}$,  the averaged output layer prediction is
(i)  $T_{c}(\varepsilon = - 2)\approx 1.3769$, and (ii)  $T_{c}(\varepsilon = + 2)\approx 3.1071$.
  These  results  of  $T_{c}(\varepsilon)$ are close to the values of $T_{c} \approx 1.4182$($T_{c} \approx 3.1201$), obtained with  Eq.   (\ref{Eq:8}).
  The discrepancy  is mainly related to  the role of unchanged energy $\Delta E = 0$ in the modified update rule when we generate the configurations, while this $\Delta E = 0$  is reasonably neglected in our calculation, see Eq. (\ref{Eq:12}).
In Table \ref{table1}, we have provided a summary of the values of $T_{c}(\varepsilon)$ that were obtained using the modified Metropolis method by means of MC simulations (literature), and supervised machine learning (our study).
 As summarized in this table, we see that  MC and ML results are almost in agrement with each other.
 This explains the best performance of ML methods that CNN  has the potential to exhibit phases and the transition temperature in unexplored out-of-equilibrium systems as well.

\begin{table}[hbpt]
\caption{\small{A summary of the values of transition temperature $T_{c}(\varepsilon)$ for $\varepsilon = \pm 2$ computed via supervised ML compared with exact result as well as the MC result reported in~\cite{Kumar2020}.  In our study, the equilibrium transition $T_{c}(\varepsilon = 0)$ is used for validation. The error estimates are given in parentheses.}} \label{table1}
 \centering
\begin{tabular}{cccc}  \hline
  Parameter           &  $T_{c}(\varepsilon)$ Exact  &   Machine Learning   & Monte Carlo   \\
  $\varepsilon $  &    (This Work) Eq. (\ref{Eq:8})  &  $T_{c}^{\texttt{ML}}$ (This Work) &  $T_{c}^{\texttt{MC}}$ Ref. \cite{Kumar2020}  \\  \hline \hline
  0   & $2/\ln(1+ \sqrt{2}) \approx2.2692 $  &$2.2687(15)$  & $-$   \\ \hline
 -2 & $ 5/4 \ln(1 + \sqrt{2}) \approx 1.4182$   &  $1.3769(87)$ &  $1.3604(3)$ \\ \hline
 +2  & $ 11/4\ln(1+ \sqrt{2}) \approx 3.1201$  & $ 3.1071(175)$ &$ 3.1267(4)$      \\ \hline \hline
\end{tabular}
\end{table}

In conclusion, CNN is easily programmable using more convoluted software libraries, and it can be advanced to identify the non-equilibrium phase transitions from typical raw lattice configurations generated by the modified Metropolis MC simulations.
Investigating whether or not this numerical method can be extended to  the non-equilibrium phase transitions in an active spherical model is one of the fascinating questions that might be asked in this area. The spherical model is another model that can be exactly solved.
 In practice, it is used to characterize a wide variety of critical phenomena, including the ferromagnetic transition and the Bose-Einstein condensation, for example.

% Remark
In this particular piece of work, we focused solely on the model's \emph{static} characteristics.
It has come to the attention of the authors that the parameter $\varepsilon$ has been included here to only play the role of violating the DBC. In a remarkable turn of works, the subsequent focus of our research will be on the mathematical formalization as well as its complete physical description. Therefore, investigating the \emph{dynamical} features of the models that violate DBC signifies a more intriguing potential course of the future direction.
%%%%%%%%%%%%%%%%%%%%%%%%%%%%%%%%%%%%%%%%%%%%%%%%%%%%%%%%%%%%%%%%%%%%%%%%%%%%%%%%%
\subsection*{Acknowledgements}
Mulugeta Bekele and DW  would like to thank International Science Programme, Uppsala, Sweden for the support not only in providing the facilities of Computational and Statistical Physics lab but also in covering all our travel as well as local expenses in visiting  Indian Institute of Science, Bangalore, India. DW would like to thank Addis Ababa University and Dire Dawa University for financial support during his research work.
%%%%%%%%%%%%%%%%%%%%%%%%%%%%%%%%%%%%%%%%%%%%%%%%%%%%%%%%%%%%%%%%%%%%%%%%%%%%%%%%%
%\section*{References}
%%%%%%%%%%%%%%%%%%%%%%%%%%%%%%%%%%%%%%%%%%%%%%%%%%%%%%%%%%%%%%%%%%%%%%%%%%%%%%%%%

%%%%%%%%%%%%%%%%%%%%%%%%%%%%%%%%%%%%%%%%%%%%%%%%%%%%%%%%%%%%%%%%%%%%%%%%%%%%%%%%%
\appendix
%%%%%%%%%%%%%%%%%%%%%%%%%%%%%%%%%%%%%%%%%%%%%%%%%%%%%%%%%%%%%%%%%%%%%%%%%%%%%%%%%
\section{Appendix (Supplementary Page) }\label{appendix1}
%%%%%%%%%%%%%%%%%%%%%%%%%%%%%%%%%%%%%%%%%%%%%%%%%%%%%%%%%%%%%%%%%%%%%%%%%%%%%%%%%
%%%%%%%%%%%%%%%%%%%%%%%%%%%%%%%%%%%%%%%%%%%%%%%%%%%%%%%%%%%%%%%%%%%%%%%%%%%%%%%%%
\subsection[\appendixname~\thesubsection]{Graphical Solution to $T_{c}(\varepsilon)$ Eq.~(\ref{Eq:8})}\label{appendix:A1}
\begin{figure}[hbpt]
     \centering
     $\begin{array}{cc}
     \includegraphics[width=0.45 \columnwidth]{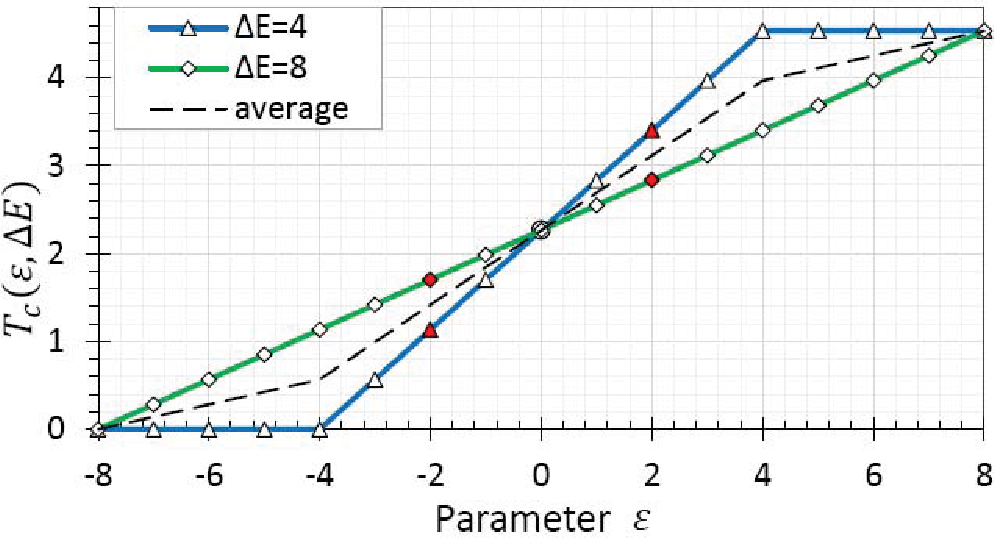} &
     \includegraphics[width=0.45 \columnwidth]{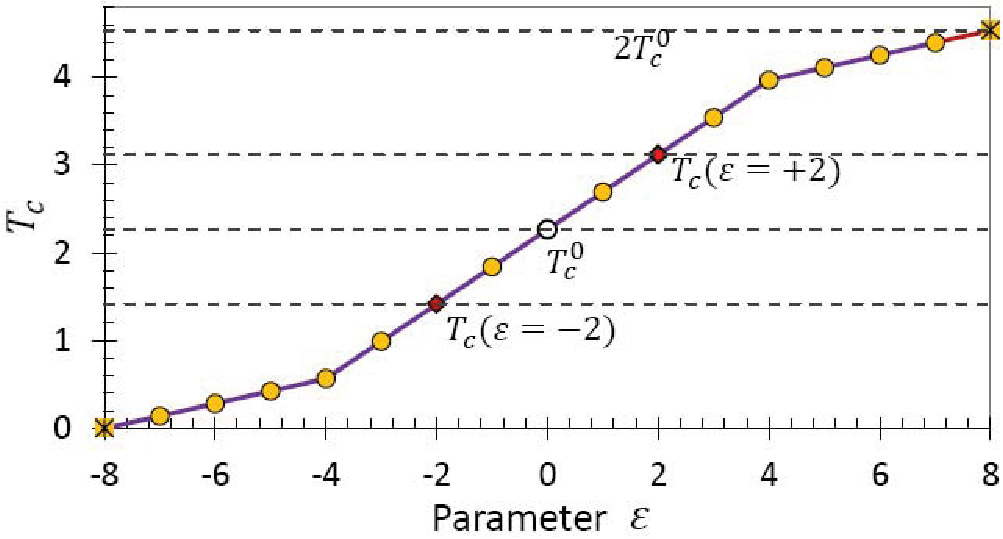}     \\
     (a) & (b)
     \end{array}$   
   \caption{\small{Critical temperature $T_{c}$  as a function of parameter $\varepsilon$. (a) Plot of Eq. (\ref{Eq:13}) with varying $\Delta E$ (see keys). The dashed line at the middle is equal to average of $T_{c}$ for two $\Delta E$ values.
(b) Plot of Eq. (\ref{Eq:8}) and this is the same as the average $T_{c}$ shown in "a".
 The horizontal dashed lines are $2T_{c}^{0}\approx 4.5384$, $ T_{c}(\varepsilon = + 2) \approx 3.1201$,  $T_{c}^{0}\approx 2.2692$  and  $ T_{c}(\varepsilon = - 2)    \approx 1.4182$.}} \label{fig:6}
\end{figure}

%%%%%%%%%%%%%%%%%%%%%%%%%%%%%%%%%%%%%%%%%%%%%%%%%%%%%%%%%%%%%%%%%%%%%%%%%%%%%%%%%
Make use of the transition rate for flipping a spin ($\sigma_{i}\rightarrow -\sigma_{i} $), Eq. (\ref{Eq:5}), and the basic definition of the energy change, $\Delta E = \{-8, -4, 0, 4, 8 \}$, it is important to consider the following two main cases:
\begin{itemize}
  \item [(i)] First one can simply verify  that the modified  algorithm~(\ref{Eq:5}) still satisfies the DBC  when $|\varepsilon| \geq  8$.
  This can be described as follows.
\begin{itemize}
  \item [a)] Assume for $\varepsilon \geq 8$ which implies that $ \varepsilon \pm \Delta E  \geq 0 $.
  Subsequently, the transition rates are
\begin{equation*}
 W(\sigma_{i}\rightarrow -\sigma_{i}) =	 \texttt{e}^{-( \Delta E + \varepsilon)/T}, \; \texttt{and} \; \;
W(-\sigma_{i}\rightarrow \sigma_{i}) =	 \texttt{e}^{-(- \Delta E + \varepsilon)/T},
\end{equation*}
where the ratio becomes
\begin{equation}\label{Eq:PhaseTransition4}
 \texttt{w}(\varepsilon \geq 8) =	 \texttt{e}^{-2 \Delta E/T }.
\end{equation}
Therefore, this satisfies the DBC; though at an effective temperature $T_{\texttt{eff}} = T/2$.
As a result, the \emph{equilibrium} transition temperature equals $T_c(\varepsilon \geq 8) = 2 T_{c}^{0}$, where $T_{c}^{0} = T_c(\varepsilon =0)$ refers to the transition temperature of this model~\cite{Onsager1944}.
\item [ b)] If we consider $\varepsilon \leq - 8$, it follows that $ \varepsilon \pm \Delta E  \leq 0 $ meaning that $ W(\sigma_{i}\rightarrow -\sigma_{i}) \equiv  W(-\sigma_{i}\rightarrow \sigma_{i})=1$, with the ratio  $\texttt{w}(\varepsilon \leq - 8)=1$.
Thus, the DBC is satisfied in this case within the limit that $T_{\texttt{eff}} \rightarrow \infty$,
indicating that there is \emph{no phase transition}~\cite{EndNote}.
\end{itemize}
 \item [(ii)] Now the second case ($|\varepsilon| < 8$) breaks the DBC since it is impossible to obtain a unique $T_{\texttt{eff}}$ in which the transition probabilities of the given $\Delta E$ can respect the DBC. We can explain this as shown below.
\begin{itemize}
  \item [a)]  Let we consider $0 <\varepsilon < 8$.  It follows that,
\begin{equation}\label{Eq:PhaseTransition6}
 \underbrace{W(\sigma_{i}\rightarrow -\sigma_{i}) =	 \texttt{e}^{-( \Delta E + \varepsilon)/T}, \; \texttt{and} \;
W(-\sigma_{i}\rightarrow \sigma_{i}) =	\texttt{MIN} \left[1,  \texttt{e}^{-(- \Delta E + \varepsilon)/T}\right]}_{\texttt{for \; } \Delta E> 0}
\end{equation}
and
\begin{equation}\label{Eq:PhaseTransition7}
\underbrace{ W(\sigma_{i}\rightarrow -\sigma_{i}) =	 \texttt{MIN} \left[1, \texttt{e}^{-( \Delta E + \varepsilon)/T}\right], \; \texttt{and} \;
W(-\sigma_{i}\rightarrow \sigma_{i}) =	  \texttt{e}^{-(- \Delta E + \varepsilon)/T}}_{\texttt{for \; } \Delta E < 0}.
\end{equation}
Here, in both (\ref{Eq:PhaseTransition6} and \ref{Eq:PhaseTransition7}), the ratio of the  transition probabilities is subject to  the value of $\Delta E$, implying that it is impossible to find a unique value of $T_{\texttt{eff}}$.
If a phase transition occurs, then the value of $T_c$  must satisfy $T_{c}^{0} < T_{c} < 2 T_{c}^{0}$.
  \item [ b)]
If we follow the same arguments for $-8 < \varepsilon < 0$, it can be inferred that  $T_c$ is expected to be in the interval $0 < T_{c} <  T_{c}^{0}$.
\end{itemize}
\end{itemize}
Recall the definition of the energy change within the \emph{equilibrium} Ising model that, $\Delta E = \{-8, -4, 0, 4, 8 \}$.
It can be noted from Ref.~\cite{Kumar2020} that,  $T_c =0$  for $\varepsilon \leq \Delta E_{\texttt{min}} = -8$, it increases from $0$ with increasing $\varepsilon$ from $-8$ to $\Delta E_{\texttt{max}} = 8$, and becomes $ 2 T_{c}^{0}$ for $\varepsilon \geq \Delta E_{\texttt{max}}$.
Explicitly,  as required for the purpose of this work, we are essentially interested in some  $\varepsilon$ values lie in $- 8 < \varepsilon < 8$, see Figure \ref{fig:6}.
For this  $\varepsilon$ values, $T_{c}(\varepsilon, \Delta E)$ can be discussed as follows.
With positive  $\Delta E = \{4, 8\}$, from Eq. (\ref{Eq:PhaseTransition6}), assume the case $W(-\sigma_{i}\rightarrow \sigma_{i})=1$ and hence the ratio becomes $ \exp[-(\Delta E +\varepsilon)/T]$.
Comparing this to that of the equilibrium case ($\varepsilon = 0$) at temperature $T^{0}$, one can obtain $T \Delta E  = T^{0}(\Delta E +\varepsilon)$.
 Similarly, for negative $\Delta E = \{-4, -8\}$, from (\ref{Eq:PhaseTransition7}) we find $ T \Delta E = T^{0}(\Delta E - \varepsilon)$.
 As a result
\begin{equation}\label{Eq:12}
   T \Delta E =
\left\{
  \begin{array}{ll}
T^{0}(\Delta E +\varepsilon), & \hbox{for \; }\Delta E > 0; \\
    T^{0}(\Delta E - \varepsilon), & \hbox{for \; } \Delta E < 0 .
  \end{array}
\right.
\end{equation}
This Eq. (\ref{Eq:12}) allows us to relate a temperature $T(\varepsilon \neq 0)$ to  $T(\varepsilon = 0)$. Since this relation provides different values for different $\Delta E$,  it is impossible to uniquely  map the probability distribution in the NESS to the equilibrium distribution at a given  $T^{0}$.
  %This confirms the nonequilibrium nature of this model for $\varepsilon \neq 0$.
To get a unique result of $T_{c}$, we need to find the average over its different values obtained by using the possible values of $|\Delta E| = \{ 4, 8\}$.
At the transition point, Eq. (\ref{Eq:12}) implies that $ T_{c} \Delta E = T_{c}^{0}(\Delta E \pm \varepsilon)$. We can write in its simple form as
 \begin{equation}\label{Eq:12}
 T_{c}(\varepsilon, \Delta E) = \left( \frac{|\Delta E|  + \varepsilon }{|\Delta E|}\right) T_{c}^{0},
 \end{equation}
where $\Delta E\neq 0$ and  $T^{0}_{c} = 2/\ln(1 + \sqrt{2})$.
 The possible values of $T_{c}$ in Eq. (\ref{Eq:12}) for $\Delta E = \{-8, -4, 4, 8\}$ are shown in Figure \ref{fig:6}(a).
In order to obtain a unique value of $T_{c}$, we need to calculate the average over the different values of $T_{c}$ that can be found from different choices of $\Delta E$\footnote{As a matter of fact only $\Delta E = \{4, 8\}$ can be used.}.
Accordingly, we can use Eq. (\ref{Eq:13}) to get the exact solution that $T_{c}^{\texttt{exact}}(\varepsilon) = (1+3\varepsilon/16)T_{c}^{0}$ (for $-4 \leq \varepsilon \leq 4$), or
\begin{subequations} \label{Eq:13}
\begin{eqnarray}
\label{Eq:13a}
 T_{c}^{\texttt{exact}}(\varepsilon) =  \frac{8 + \varepsilon }{8\ln(1+\sqrt{2})}, & \texttt{for} & -8 \leq \varepsilon <-4; \\
 \label{Eq:13b}
T_{c}^{\texttt{exact}}(\varepsilon) =    \frac{16 +3 \varepsilon }{8\ln(1+\sqrt{2})}, & \texttt{for}& -4 \leq \varepsilon \leq 4; \\ \label{Eq:13c}
T_{c}^{\texttt{exact}}(\varepsilon) =  \frac{24 + \varepsilon }{8\ln(1+\sqrt{2})}, &  \texttt{for} & 4 < \varepsilon \leq 8 .
\end{eqnarray}
\end{subequations}
Alternatively, one can use $t_{c} = T_{c}/T_{c}^{0}$ to rewrite as
\begin{equation}\label{Eq:17}
\begin{array}{cc}
 \left.\begin{array}{c}
   \includegraphics[width=0.3 \columnwidth]{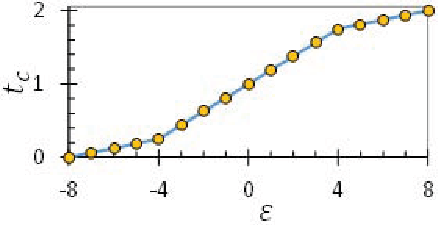}
\end{array}\right\} &
\begin{array}{l}
   t_{c} =  \left\{
      \begin{array}{ll}
         0.5 + \varepsilon/16, & \hbox{for} -8 \leq \varepsilon <-4; \\
        1 + 3\varepsilon/16, & \hbox{for} -4 \leq \varepsilon \leq 4;\\
        1.5 + \varepsilon/16, & \hbox{for}  4 < \varepsilon \leq 8.
      \end{array}
    \right.
\end{array}
\end{array}
\end{equation}
Implicitly, it can be inferred from Figure \ref{fig:6}(b) that  Eq. (\ref{Eq:13b}) is efficient to discuss the non-equilibrium phase transition, where $T_{c}(\varepsilon = 0)=T_{c}^{0}$ and  $t_{c}^{0}= 1$.
%%%%%%%%%%%%%%%%%%%%%%%%%%%%%%%%%%%%%%%%%%%%%%%%%%%%%%%%%%%%%%%%%%%%%%%%%%%%%%%%%
\subsection{Schematic Representation of the \texttt{Modified}  Metropolis MC Simulation}\label{appendix:A2}
%%%%%%%%%%%%%%%%%%%%%%%%%%%%%%%%%%%%%%%%%%%%%%%%%%%%%%%%%%%%%%%%%%%%%%%%%%%%%%%%%
\begin{figure}[hbpt]
  \centering
\includegraphics[width=0.5\columnwidth]{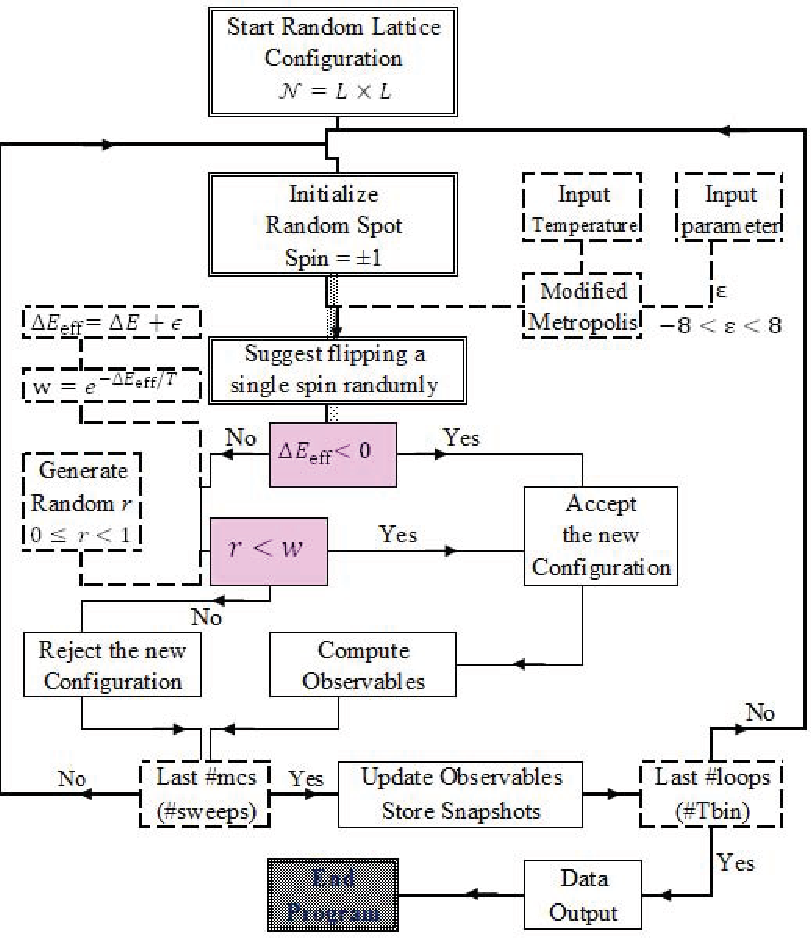}\\
  \caption{\small{An schematic representation of the \texttt{modified}  Metropolis MC simulation ($\Delta E_\texttt{eff} = \Delta E + \varepsilon$).
First initialize a random configuration of $L\times L$ spins, then randomly choose a spin site to flip. Next compute $\Delta E_{\texttt{eff}}$ (\ref{Eq:dEeff}) where $\Delta E$ is readily from definition, if  $\Delta E_{\texttt{eff}} < 0$  accept the flip, otherwise  accept the flip with probability $\texttt{w}=\texttt{e}^{- \Delta E_{\texttt{eff}}/T}$. This is numerically implemented by generating a \texttt{random} number $r=[0,1)$, if $r < \texttt{w}$  accept the flip and  reject otherwise.
We perform a sweep over the entire lattice of $\mathcal{N}=L\times L$ spins 10 times, such that there is a total number of $10\mathcal{N}$ possible spin flips to improve the generation of steady state data.
Note that we recover the original Metropolis when $\varepsilon=0$.
}}\label{fig:9}
\end{figure}.
%%%%%%%%%%%%%%%%%%%%%%%%%%%%%%%%%%%%%%%%%%%%%%%%%%%%%%%%%%%%%%%%%%%%%%%%%%%%%%%%%\newpage
\subsubsection*{Representative Spin Configurations}
\begin{figure}[hbpt]
     \centering
 $ \begin{array}{cc}
    \underbrace{\begin{array}{c}
     \boxed{\includegraphics[width=0.275 \columnwidth]{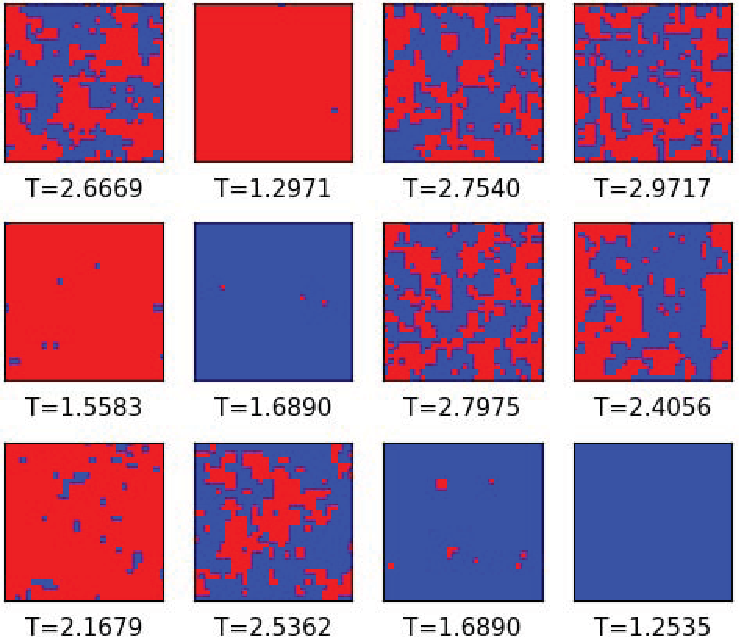}}    \\
   \texttt{(a) } \varepsilon = 0    \end{array}}
     & \underbrace{\begin{array}{cc}
      \boxed{ \includegraphics[width=0.275 \columnwidth]{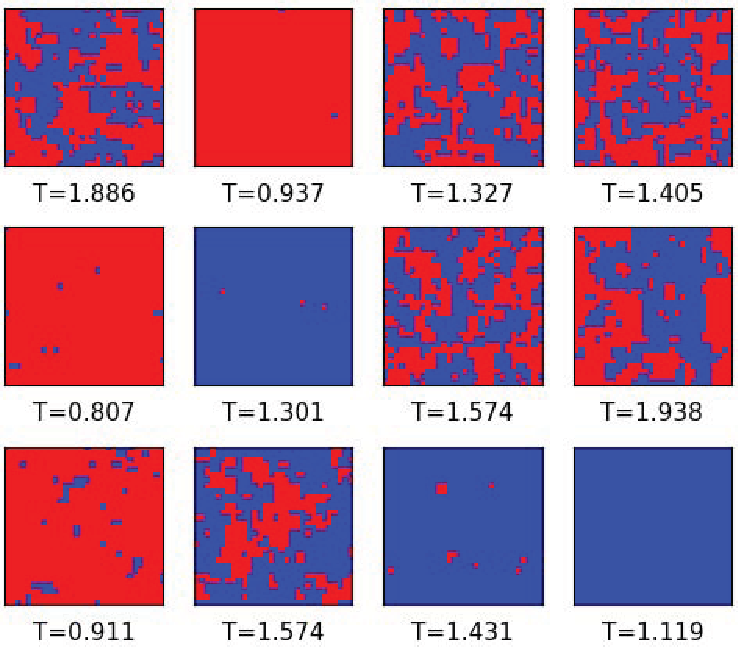}}
           &       \boxed{ \includegraphics[width=0.275\columnwidth]{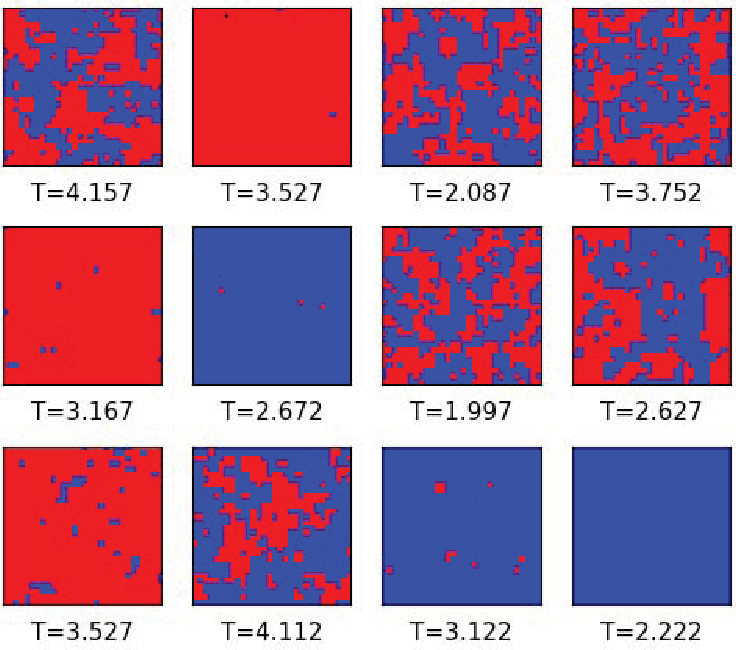}}  \\
       \texttt{(b) } \varepsilon=-2  &  \texttt{(c) } \varepsilon = +2
      \end{array} }\\
     \texttt{Equilibrium Representative}  &  \texttt{Non-equilibrium Representative} \\
     (\texttt{Train Data-set})   &  (\texttt{Test Data-sets})
  \end{array}$
 \caption{\small{Representative  spin  configurations ($30 \times 30$) from the training  (panel a),  and test (panels  b and c) datasets. The low temperature ($T< T_{c}(\varepsilon)$) configurations tend to be predominately aligned in either the "down" (red) or "up" (blue) directions.}} \label{fig:samples}
\end{figure}

%%%%%%%%%%%%%%%%%%%%%%%%%%%%%%%%%%%%%%%%%%%%%%%%%%%%%%%%%%%%%%%%%%%%%%%%%%%%%%%%%
Figure \ref{fig:samples} demonstrates a  system size of  $ 30 \times 30$  representative Ising spin  configurations at various values of temperature. There are 12 samples from the training (a)  and 24 samples are from test datasets (b and c).
 %We use these configurations as they  play the role of \emph{2D images} to train and test the neural network.
%%%%%%%%%%%%%%%%%%%%%%%%%%%%%%%%%%%%%%%%%%%%%%%%%%%%%%%%%%%%%%%%%%%%%%%%%%%\newpage
\subsection{Architecture of Convolutional Neural Network}\label{appendix:A3}
\begin{figure}[hbpt]
     \centering
\includegraphics[width=1\columnwidth]{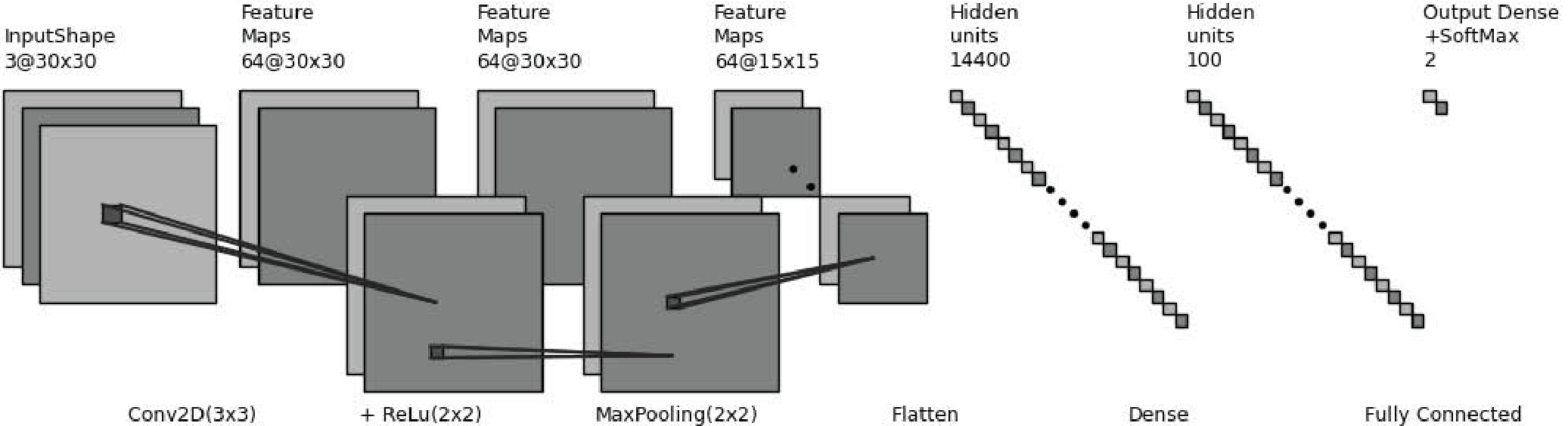}\\ % or we use L30bin in practice
\small{(a) A simple CNN model: A neural network model constructed from a convolution layer and a fully connected layer.}
$\begin{array}{cc}
 \includegraphics[width=0.35\columnwidth]{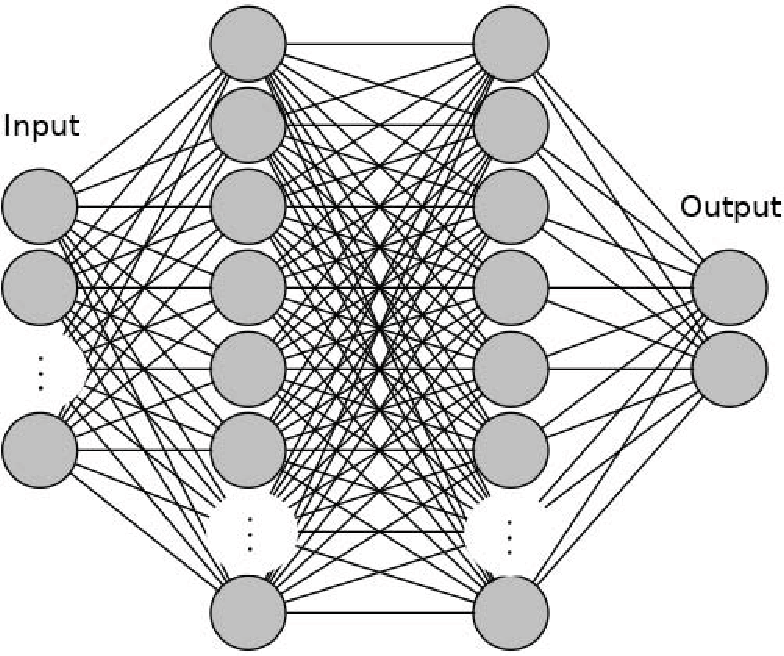}
 &
\includegraphics[width=0.375\columnwidth]{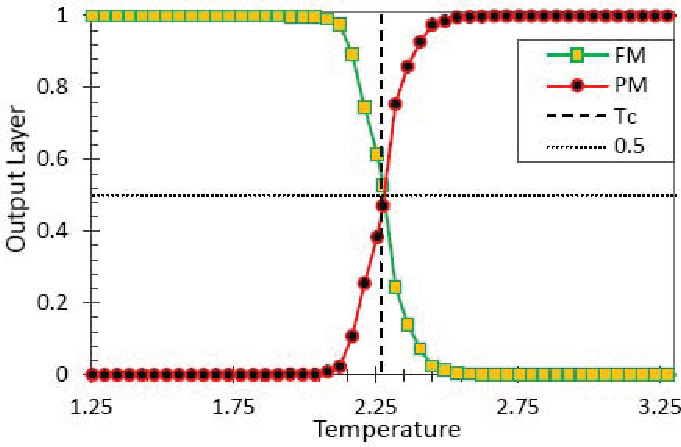} \\
\small{\texttt{(b) Fully connected  layer}}  &\small{\texttt{(c) A plot of prediction versus temperature~\cite{Carrasquilla2017}.}}
\end{array}$
\caption{\small{Schematic diagram  of the machine learning architecture~\cite{ConvNet2017}(an example).}}
    \label{fig:cnn1}
\end{figure}

%%%%%%%%%%%%%%%%%%%%%%%%%%%%%%%%%%%%%%%%%%%%%%%%%%%%%%%%%%%%%%%%%%%%%%%%%%%%%%%%%
We build a simple convolutional neural network (CNN), implemented with TensorFlow \cite{TensorFlow2015} $\texttt{keras}$ sequential model, to perform supervised machine learning (ML) on the Ising spin configurations sampled by the (effective) Metropolis MC simulation. An example of the architecture~\cite{ConvNet2017} of our model is represented as illustrated in Figure~\ref{fig:cnn1}.
As shown in this example, the input layer consists of a square lattice ($L \times L$) Ising spin configurations.
The input-shape specified on the input layer represents the shape of our input data (i.e., snapshot images). Here, the example with $L=30$ shows that  each image is  30 pixels wide and 30 pixels high, and has three ($\texttt{RGB}$) color channels which gives us an  $\texttt{Inputshape} = (30, 30, 3)$. In this model, the first hidden layer is a two dimensional convolutional layer ($\texttt{Conv2D}\_1$). This layer has 64 output filters (each of $3\times 3$ kernel size) with a single stride  ($\texttt{Stride} = 1$), and we use rectified linear unit ($\texttt{ReLU}$) activation function. In addition, specifying $\texttt{padding}$ and  enabling the valid $\texttt{periodic-padding}$ helps to account periodic boundary conditions.  Note that  the choice for the kernel size of $3 \times 3$ is generally a very common size to use, but the chosen number of output filters specified is arbitrary. One can choose different values of these parameters through observations during training the model of interest.

Next we add a second $\texttt{Conv2D}\_2$ with the same specs as the first $\texttt{Conv2D}\_1$.
Similar to $\texttt{Conv2D}\_1$, this $\texttt{Conv2D}\_2$ has also 64 filters.
Note also the choice of 64 here is arbitrary, even though having more filters in the later layers than in earlier layers is usually recommended in some cases.
Optionally, one can add a max-pooling layer ($\texttt{MaxPool2D}$) to pool and reduce the dimensionality of the data\footnote{We use Ref.~\cite{Ethem2004} for the fundamental understanding of max-pooling, padding, convolutional filters, and CNN }.
Finally, we need to flatten the output from the convolutional layer and pass it to a $\texttt{Dense}$ layer.
We use an appropriate number of epochs (e. g., $\texttt{epoch} = 4$) and  apply a $\texttt{Dropout}$ regularization in the $\texttt{Dense}$ layer in order to avoid overfitting~\cite{hinton2012dropout}.
In our case, the last $\texttt{Dense}$ layer has two nodes ($\texttt{Dense} = 2$) which means that one for each classes; namely FM and PM states.
 In addition, we use the $\texttt{Softmax}$ activation function on the last $\texttt{Dense}$ layer so that the output for each sample is a probability distribution over the outputs of each classes.

As an example, Figure \ref{fig:cnn1}(c) shows the \emph{prediction} (by prediction we mean the average output values of the final $\texttt{Dense}$ layer) for  configurations of different temperatures $T$ where $\varepsilon =0$ was used here.
 The red ($\bullet$)  and the green ($\Box$)  curves represent the average prediction of the FM and PM phases, respectively. Here, the sum of the two prediction should be $P_{\texttt{FM}}+P_{\texttt{PM}}=1$.
The temperature at which the two curves intersect indicates the temperature at which CNN switches between classifying configurations as `FM' versus `PM' phases. The crossing point is also known as point of maximal confusion~(POM).
The horizontal dashed line represents an estimate of prediction $P =0.5$, while the vertical dashed line indicates  the model's crossing temperature $T^{*}$.
(Note that these notations are also same for detail results presented in Sec. \ref{section:Results}.)
Remarkably, the value of  $T^{*}$ agrees with the exact  result, $T^{*} \approx  T_{c}^{0}$.
For the detailed numerical analysis presented in section \ref{section:Results}, we stick ourselves to  $\varepsilon = \pm 2$. The example of qualitative dependence of the critical temperature $T_{c}$ on $\varepsilon$ is presented in \ref{appendix:A4}.
To this end, a basic understanding of PT  between the PM phase ($T> T_{c}$) and the FM phase ($T< T_{c}$), permits and helps our efforts to categorize the two different types of configurations via ML~\cite{Carrasquilla2017}.
%%%%%%%%%%%%%%%%%%%%%%%%%%%%%%%%%%%%%%%%%%%%%%%%%%%%%%%%%%%%%%%%%%%%%%%%%%%%%%%%%
\subsection{Qualitative Dependence of $T_c$ on the Parameter $\varepsilon$} \label{appendix:A4}
\begin{figure}[hbpt]
     \centering
  $\begin{array}{ll}   
 \texttt{(a) } &   \texttt{(b) } \\
\includegraphics[width=0.55 \columnwidth]{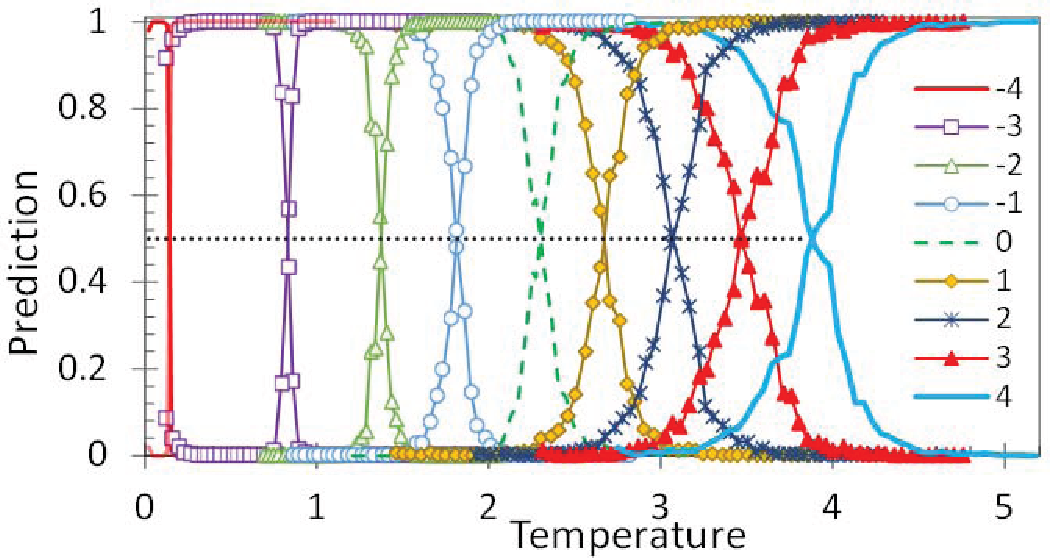} &
\includegraphics[width=0.4 \columnwidth]{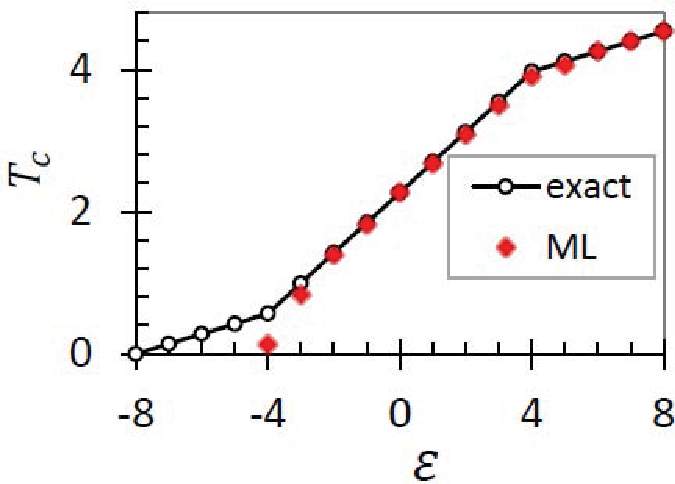}
\end{array}$
\caption{\small{(a) Prediction versus temperature for various $\varepsilon$ values. (b) Qualitative dependence of $T_{c}$ on $\varepsilon$ where we compare the numerical result of $T_{c}^{ \texttt{ML}}$ with that of the exact result $T_{c}^{ \texttt{exact}}$.}} \label{fig:7} %The equilibrium ($\varepsilon=0$) is also shown for reference.
\end{figure}

%%%%%%%%%%%%%%%%%%%%%%%%%%%%%%%%%%%%%%%%%%%%%%%%%%%%%%%%%%%%%%%%%%%%%%%%%%%%%%%%%
Figure \ref{fig:7}
shows the plots of prediction versus  $T$ for $L= 30$ and for some values of $\varepsilon$ to show the dependence of the critical temperature $T_c$ on $\varepsilon$.
The positions at which the curves are crossing each others give estimates of $T_c(\varepsilon)$.
Notice the shifting of $T_c$ to higher values with increasing $\varepsilon$.
The inset of this figure shows a plot of $T_c(\varepsilon)$, estimated as the value of $T$ at which curves cross, as a function
of $\varepsilon$. For $\varepsilon =0$, the model has an equilibrium phase transition at $T_c(\varepsilon=0)\approx2.2692$.
It is clear from the plot in the inset that $T_c$ approaches zero for large negative values of $\varepsilon$ and it is $\simeq 2T_c(\varepsilon=0)$ for $\varepsilon=8$.
This is in agreement with Eqs. (\ref{Eq:6}) and  (\ref{Eq:8}) for  $-4 < \varepsilon < 8$.
However, the model fails to detect the transition temperature for $-8 < \varepsilon < -4$.
%%%%%%%%%%%%%%%%%%%%%%%%%%%%%%%%%%%%%%%%%%%%%%%%%%%%%%%%%%%%%%%%%%%%%%%%%%%%%%%%%  
\subsection{FSS of the Transition Temperature and the Critical Exponent~($\gamma$) }\label{appendix:A6}
\begin{figure}[hbpt]
     \centering
    $  \begin{array}{ccc}
     \texttt{(a) } T_{c}^{0} = \frac{2}{\ln(1+ \sqrt{2})} &
     \texttt{(b) }  T_{c}(\varepsilon = -2) = \frac{5}{ 4 \ln(1 + \sqrt{2})} &
     \texttt{(c) }  T_{c}(\varepsilon = +2) = \frac{11}{ 4 \ln(1 + \sqrt{2})} \\
                 \includegraphics[width=0.3 \columnwidth]{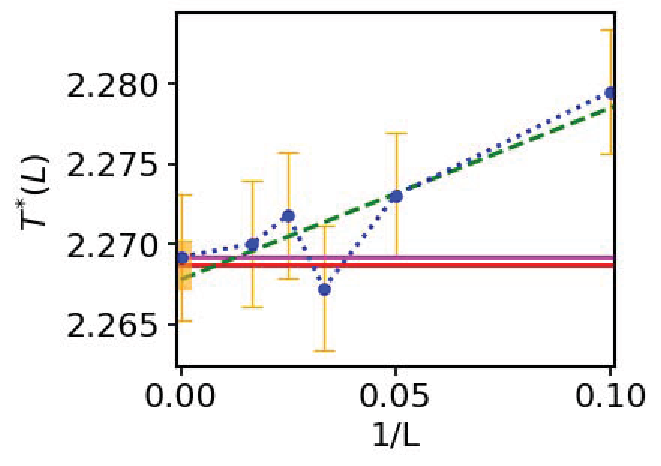} &
                  \includegraphics[width=0.3 \columnwidth]{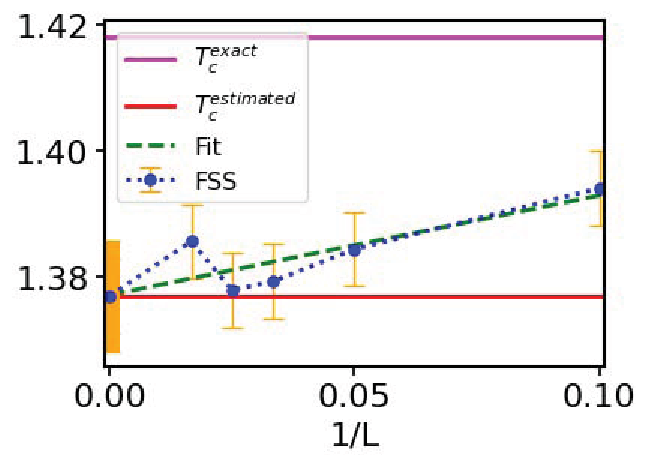} &
                 \includegraphics[width=0.3 \columnwidth]{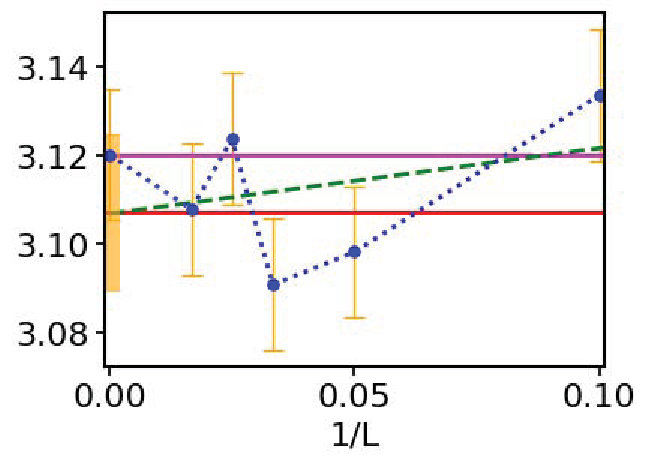}
               \end{array} $
\caption{\small{The crossing temperature $T^{*}(L)$ as a function of  1/L. The horizontal red line (see keys) refers to the numerical $T_{c}^{\textit{ML}}$ (Table \ref{table1}), and the magenta line represents 
   $T_{c}^{\textit{exact}}$ (Eq. \ref{Eq:13b}) as shown in each `a', `b' and `c'. }} \label{fig:FSSTstar}
\end{figure}

%%%%%%%%%%%%%%%%%%%%%%%%%%%%%%%%%%%%%%%%%%%%%%%%%%%%%%%%%%%%%%%%%%%%%%%%%%%%%%%%%
Figure \ref{fig:FSSTstar} demonstrates the FSS analysis of the crossing temperature $T^{*}(L)$  as a function of $1/L$ for $L= \{ 10, 20, 30, 40, 60 \}$.
 The horizontal red line (see keys) refers to the numerical $T_{c}^{\textit{ML}}$ (Table \ref{table1}), and the  magenta line represents the critical temperature in thermodynamic limit that is calculated using Eq. (\ref{Eq:13b}),  $T_{c}(\varepsilon = \pm 2) = (16 + 3\varepsilon)/8 \ln(1 + \sqrt{2})$,  where the known result of $T_{c}(\varepsilon=0)$ is shown for reference. The size of error  bars is equal to one standard deviation statistical uncertainty \footnote{See the supplementary material  in Ref. \cite{Carrasquilla2017}}.
 The numerical results are (a) $T_{c}^{\textit{ML}} \simeq 2.2687(15)$, (b) $T_{c}^{\textit{ML}} \simeq 1.3769(87) $ and  (c) $T_{c}^{\textit{ML}} \simeq 3.1071(175)$.

\begin{figure}[hbpt]
     \centering
\includegraphics[width=1 \columnwidth]{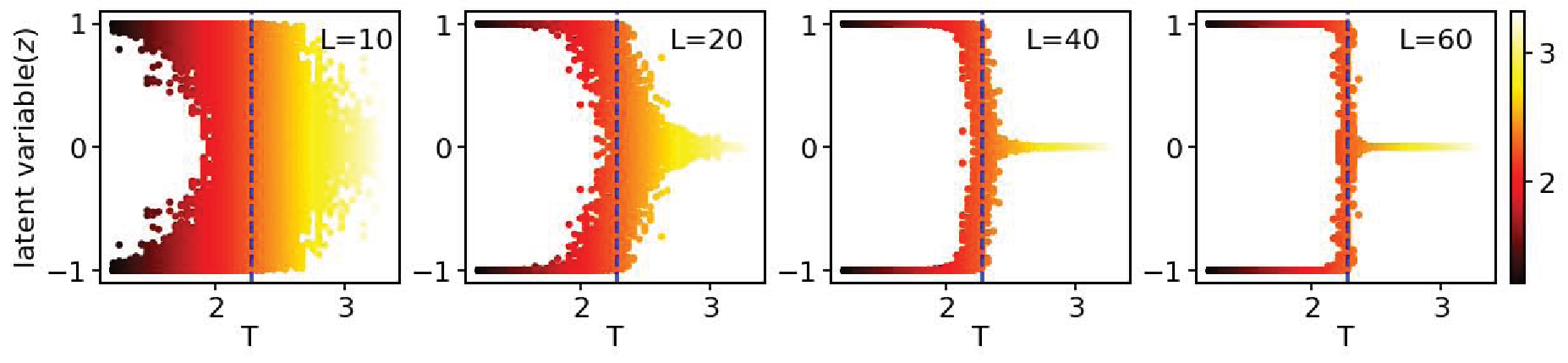}
   \caption{\small{Scatter plots of the latent variable $z$ versus $T$ for four systems of each linear size $L$ shown at the upper right. This is the case of equilibrium ($\varepsilon =0$) model. The blue dashed line within each system denotes $T_{c}^{0}=2/\ln(1+\sqrt{2})$. A gradient color at the right panel help to illustrates the temperature $T$ and it reflects the nature of the phase diagram as we move from low $T$ ($T<T_{c}^{0}$) to high $T$ ($T>T_{c}^{0}$).}} \label{fig:z0}
\end{figure}

%%%%%%%%%%%%%%%%%%%%%%%%%%%%%%%%%%%%%%%%%%%%%%%%%%%%%%%%%%%%%%%%%%%%%%%%%%%%%%%%%
%\subsubsection*{Remark on Critical Exponent~($\gamma = 7/4$)}
%%%%%%%%%%%%%%%%%%%%%%%%%%%%%%%%%%%%%%%%%%%%%%%%%%%%%%%%%%%%%%%%%%%%%%%%%%%%%%%%%
The estimation of the critical exponent~($\gamma$) can be performed using the FSS theory $ \chi \propto (T/T_c -1 )^{-\gamma}$ where $\chi$ represents the latent susceptibility~\cite{Alexandrou2020}.
If $\tilde{z}$ denotes average absolute latent variable ($\tilde{z} = \langle |z|\rangle$), one can calculate $\chi$ as
\begin{equation}\label{Eq:susceptibility}
    \chi = \frac{\mathcal{N} \left(\left\langle \tilde{z}^{2}\right\rangle -  \left\langle \tilde{z} \right\rangle^{2}\right) }{T},
\end{equation}
recalling  $\mathcal{N} = L \times L$ and defining
$ \tilde{z} = \frac{1}{M} \sum_{k=1}^{M} |z_{k}|,$
where $M$ is the total number of configurations and $k= \{1, \cdots , M \}$.
Figure \ref{fig:z0} demonstrates the latent variable $z$  as a function of $T$ for the given configurations.
As a result, one can estimate $\gamma$ using the result of Eq.~\ref{Eq:susceptibility}  where $\tilde{z}$ can be obtained from data presented in Figure \ref{fig:z0} by  plotting $\chi L^{-\gamma/\nu}$ versus $(T-T_{c}^{0})L^{1/\nu}$.

\begin{figure}[hbpt]
     \centering
     $\begin{array}{l}
       \texttt{(a) } \varepsilon = - 2 \\
      \includegraphics[width=1.02 \columnwidth]{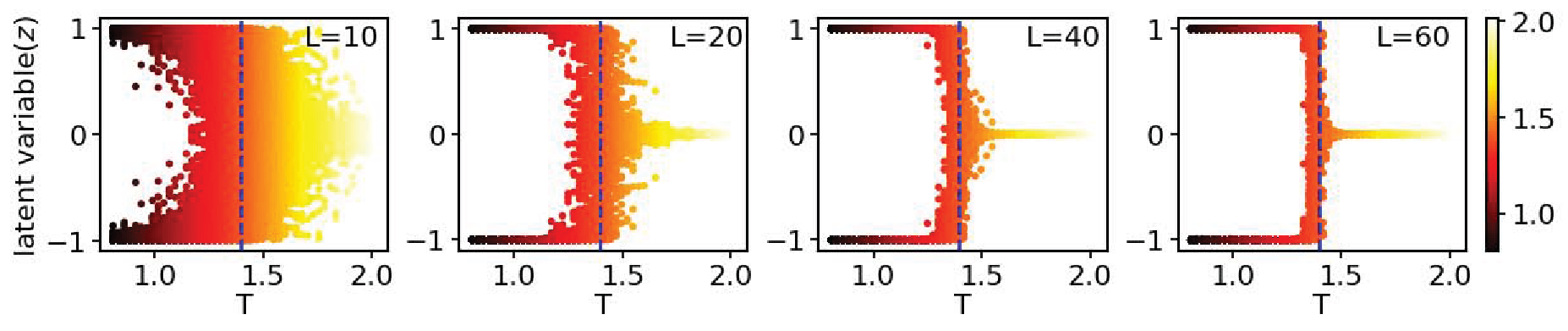}\\
      \texttt{(b) } \varepsilon = + 2 \\
       \includegraphics[width=1 \columnwidth]{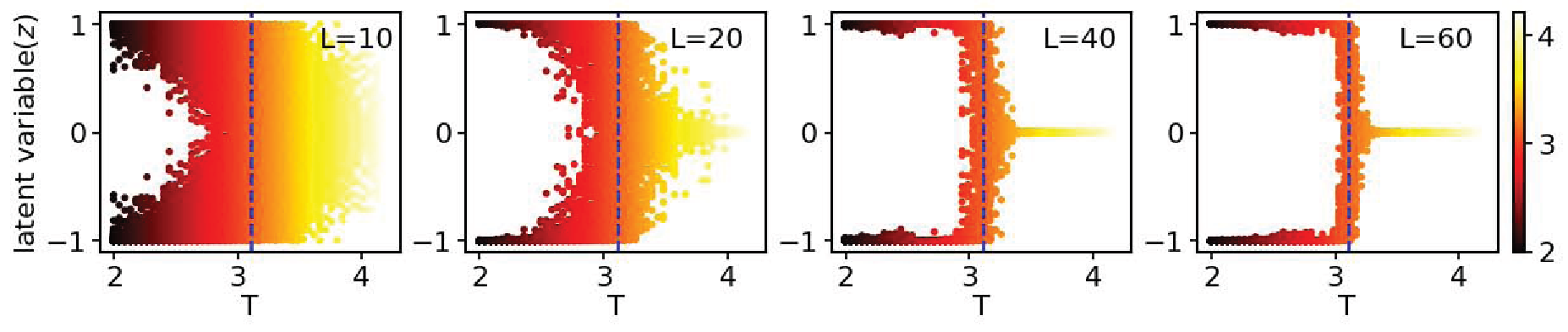}
     \end{array}$
   \caption{\small{Scatter plots of the latent variable $z$ versus $T$ for the given linear size, $\varepsilon = - 2$ (upper panel) and $\varepsilon = + 2$ (lower panel). Here the blue dashed lines denote (a) $T_{c}(\varepsilon = - 2)=(5/4)/\ln(1+\sqrt{2})$ and (b) $T_{c}(\varepsilon = - 2)=(11/4)/\ln(1+\sqrt{2})$.  The gradient color shown for each panel demonstrates the nature of the phase diagram from the low $T$ ($T<T_{c}$) to the high $T$ ($T>T_{c}$).}} \label{fig:zpm2}
\end{figure}

Therefore, it is straightforward to apply this method to the nonequilibrium case ($\varepsilon \neq 0$).
Figure \ref{fig:zpm2} shows the latent variable $z$  as a function of $T$  (a) $\varepsilon = - 2$  and (b) $\varepsilon = + 2$. 
In fact, a comprehensive investigation of this topic is underway\cite{DTetal2023}. 
%%%%%%%%%%%%%%%%%%%%%%%%%%%%%%%%%%%%%%%%%%%%%%%%%%%%%%%%%%%%%%%%%%%%%%%%%%%%%%%%%
\subsection*{Abbreviations}
\begin{tabular}{@{}ll}
CNN & Convolutional Neural Networks \\
DBC & Detailed Balance Condition    \\
FM & Ferromagnetic \\
FSS  &Finite Ssize Scaling  \\
MC & Monte Carlo \\
ML &  Machine learning \\
NESS & Non-equilibrium Steady States \\
PM & Paramagnetic  \\
PT & 	Phase Transitions \\
\end{tabular}
%%%%%%%%%%%%%%%%%%%%%%%%%%%%%%%%%%%%%%%%%%%%%%%%%%%%%%%%%%%%%%%%%%%%%%%%%%%%%%%%%
\subsection*{Data Availability}
%%%%%%%%%%%%%%%%%%%%%%%%%%%%%%%%%%%%%%%%%%%%%%%%%%%%%%%%%%%%%%%%%%%%%%%%%%%%%%%%%
The neural network result was calculated using TensorFlow~\cite{TensorFlow2015} integrated with Keras environment. Required data-sets are available from the authors for a reasonable request.
\end{document}